\documentstyle[aps,prl,preprint,floats,epsf]{revtex}

\input psfig
\long\def\simplex#1#2#3#4#5{
\begin{figure}[#1]
   \begin{center}
   \quad\\[-3cm]
   \quad
   \hbox{
   \quad 
   \parbox[t]{\hsize}{ \quad\hskip3cm\psfig{figure=#2,width=#5} 
   \vskip-0cm
   \caption[]{\small  \label{fig:#3} #4 }}
   }
   \quad
   \end{center} 
\end{figure}
}

\begin{document}

\preprint{\tighten\vbox{\hbox{\hfil CLNS 96/1450}
                        \hbox{\hfil CLEO 96-21}
}}

\title{Tau neutrino helicity from $h^{\pm}$ energy correlations}

\author{CLEO Collaboration}
\date{\today}

\maketitle
\tighten

\begin{abstract} 
We report a measurement of the magnitude of the tau neutrino
helicity from tau-pair 
events taken with the CLEO detector
at the CESR electron-positron storage ring. 
Events in which each tau undergoes the decay 
$\tau$$\rightarrow$$h\nu$, with
$h$ a charged pion or kaon,
are analyzed for energy correlations
between the daughter hadrons, yielding 
$\vert\xi_{h}\vert$ 
= $1.03\pm 0.06\pm 0.04$,
with the first error statistical and the second systematic.
\end{abstract}

\newpage

{
\renewcommand{\thefootnote}{\fnsymbol{footnote}}


\begin{center}
T.~E.~Coan,$^{1}$ V.~Fadeyev,$^{1}$ I.~Korolkov,$^{1}$
Y.~Maravin,$^{1}$ I.~Narsky,$^{1}$ V.~Shelkov,$^{1}$
J.~Staeck,$^{1}$ R.~Stroynowski,$^{1}$ I.~Volobouev,$^{1}$
J.~Ye,$^{1}$
% ...
\end{center}
 
\small
\begin{center}
$^{1}${Southern Methodist University, Dallas, Texas 75275}\\
% ...
\end{center}
 
\setcounter{footnote}{0}
}
\newpage

In the Standard Model of 
the weak interaction the tau is a
sequential lepton with a purely left-handed
weak isodoublet partner, $\nu_{\tau}$;
i.e., the tau neutrino has its spin anti-parallel to its momentum.
For the decay of the tau to a single pseudoscalar
hadron $h$ (with $h$~$=$~$\pi$ or $K$), 
the parent tau
spin is maximally correlated  
with the daughter hadron direction~\cite{tsai}.
At high beam energies, the {\it s-}channel interaction
$e^{+}e^{-}$$\rightarrow$$\gamma^{*}$$\rightarrow$
$\tau^{+}\tau^{-}$$\rightarrow$($h^{+}\bar{\nu}$)($h^{-}\nu$)
tends to have the tau spins aligned, leading to
a correlation between the angles of the 
hadron
momentum vectors as measured in the 
tau rest frames.
The Lorentz boost to the laboratory frame
also maps these angles into the observed
particle energies, 
transforming the spin-direction correlation into
an energy-energy correlation between the hadrons.
In this Brief Report we 
describe a measurement of 
the magnitude of the tau neutrino helicity
based on pseudoscalar energy-energy correlations at
a center-of-mass energy of 
$\sqrt{s}$$=$$2E_{b}$$\simeq$10.6~{\rm GeV}.  
This study complements
other recent measurements that involve other modes, 
other techniques,
and/or higher beam energies~\cite{ARGUS1,ARGUS2,ALEPH,L3,OPAL}.

The helicity, denoted here as
$\xi_{h}$ and elsewhere as $h_{\nu_{\tau}}$, is related to
the parameter $\gamma_{av}$
and the charged current couplings $g_{a}$ and $g_{v}$ 
by~\cite{kuhn,nelson,fetscher}

\begin{equation}
\xi_{h} = h_{\nu_{\tau}} = -\gamma_{av} = 
-2g_{a}g_{v}/(g^{2}_{a} + g^{2}_{v})
\end{equation}

If we define 
$c_{\pm}$ as the cosine of the angle of the
$h^{\pm}$ momentum in the parent rest frame
with respect to the boost direction,
the correlation is given by ~\cite{kuhn}
\begin{equation}
\frac{d^{2}\sigma}{dc_{+}dc_{-}} \propto 1 +  
	\frac{(2E_{b}^{2} - m_{\tau}^{2})}{(2E_{b}^{2} + m_{\tau}^{2})}
	\xi_{h}^{2}c_{+}c_{-}~.
\end{equation}
At $\sqrt{s}$$=$10.6 GeV, the net spin alignment is 
roughly 90\%.
Using instead the laboratory variables
$x_{\pm} = \sqrt{ p_{\pm}^{2} + m^{2}_{\pi} } / E_{b}$
gives the energy-energy correlation ~\cite{nelson,fetscher}
\begin{equation}
\frac{d^{2}\sigma}{dx_{+}dx_{-}} = F_{1}(x_{+},x_{-}) + 
           \xi_{h}^{2} \cdot F{_2}(x_{+},x_{-}),
\end{equation}
with $F_{1,2}$ being known kinematic functions.
This correlation is maximal and identical for a left-handed
($\xi_{h}$=$-1$) and a right-handed ($\xi_{h}$=+1) tau neutrino, and
it vanishes for no preferred handedness ($\xi_{h}$=0). It is not
explicitly parity-violating but is a consequence of parity violation
in tau decay.

The data analyzed here correspond to a luminosity of 1.64~fb$^{-1}$ 
(1.5$\times$10$^{6}$ $\tau^{+}\tau^{-}$ events),
collected at the Cornell Electron Storage Ring (CESR)
with the CLEO~II detector~\cite{CleoII}. 
A set of three concentric drift chambers in a 1.5~T axial 
magnetic field measures 
charged particle momenta with resolution 
$\sigma_{p}/p\ (\%)\simeq \sqrt{(0.15p)^2 + (0.5)^2}$, $p$ in GeV/$c$. 
Surrounding the drift chambers, but inside the superconducting
magnet coil, is a CsI(Tl) crystal electromagnetic calorimeter.
Barrel crystals surround the tracking chambers, covering 
$\vert\cos\theta\vert$$<$0.82,
with $\theta$ the angle with respect to the $e^{+}$ beam direction.
Two identical endcaps occupy 0.80$<$$\vert\cos\theta\vert$$<$0.98.
Particle time-of-flight is provided by 5~cm-thick scintillation 
counters located just inside the calorimeter in the barrel and endcap.
Muons are identified by their penetration through the calorimeter, 
coil and one or more of three 36~cm-thick slabs of magnet iron; three layers 
of Iarocci tube chambers instrument the gap behind each slab. 
A three stage hardware trigger~\cite{trigger} uses combinations of
calorimeter, tracking chamber, and time-of-flight information
to initiate detector readout.

To measure $\vert\xi_{h}\vert$ we first extract 
$\tau^+\tau^-$ $\rightarrow$ $(h^{+}\overline{\nu})(h^{-}\nu)$ 
events that are relatively free of backgrounds from non-tau
processes.  
Monte Carlo samples of events from $V$ and $V$$\pm$$A$ processes
are then generated and mixed appropriately to obtain a desired value of
$\xi_{h}$.  We determine what value of $\xi_{h}$ best fits the data
using likelihood and $\chi^{2}$ methods, 
investigating both  
two-dimensional $(x_{+},x_{-})$ 
and one-dimensional ($c_{+}c_{-}$) distributions.
The techniques are then analyzed for systematic biases and uncertainties.

Events are 
selected that have exactly two 
oppositely charged tracks 
in the fiducial volume of the tracking system
with $\vert{\rm cos}\theta\vert$$<$$0.7$
and $x_{\pm}$$>$$0.2$ and 
which project back to the $e^{+}e^{-}$ luminous region.
To reduce the contamination of non-$\tau\tau$ QED events, 
we demand that at most one track have $x_{\pm}$$>$$0.8$, 
that neither track have $x_{\pm}$$>$$0.95$, 
and that the total shower energy in the
calorimeter be less than $0.85 \sqrt{s}$.  
To remove backgrounds from final states
that involve electrons, each track must have its
calorimeter energy
less than 85\% of its measured momentum.
These criteria reduce the number of events to about 115,000.

Final states that contain photons from neutral pions are suppressed
by demanding that there be no photon-like showers in the calorimeter
that are not matched to charged tracks.
A combination of energy and isolation criteria is used to
identify such showers.
This requirement also greatly reduces the contamination from
events with a single photon, such as from the process
$e^{+}e^{-}$$\rightarrow$$\rho\gamma$. 
To suppress these radiative events further 
the two charged tracks are required to 
have an opening angle of
greater than 90$^{\circ}$ in the $r-\phi$ projection.
To minimize possible systematic
effects,
events must satisfy trigger criteria that are suitable
for two-track events and that are uniform throughout the
dataset.

The most effective variable for removing events
from $\gamma\gamma$ interactions 
is $\Theta_{\rm{min}}$, defined by
\begin{equation}
{\rm sin}\Theta_{{\rm min}} =
 \frac{\vert\vec p_\perp \vert}{E_{b} \cdot (2 - x_{-} - x_{+}) },
\end{equation}
i.e., the ratio of $\vert\vec p_\perp \vert$,
the missing net momentum transverse to the beams,
to the missing energy of the event~\cite{MasuiThesis,BKHPRD}.
This requirement also effectively suppresses 
radiative QED events in which
radiated photons go undetected.
Events are retained if ${\rm sin}\Theta_{{\rm min}}$$>$0.10.
Because particles from $\gamma\gamma$ interactions tend to have
low momentum, at least one of the tracks is also required to have 
$x_{\pm}$$>$0.3.  

Discrimination
against muons involves four mutually exclusive criteria, three of which
use information from the muon chambers and 
one of which demands the particle leave an energy deposition
in the calorimeter that is inconsistent
in magnitude and shape with that of a muon.  This last
criterion is invoked only if the charged
track projects into uninstrumented regions at the
azimuthal boundaries of the iron absorber.  Each track must pass
one of these four criteria for the event to be accepted.  
The efficiency
of the combination of these criteria to select a track as a 
pion has been measured 
using an independent subsample of hadrons in $\tau^{+}\tau^{-}$
decays involving a lepton recoiling against 
one or more neutral pions 
and a single track that is  
assumed to be a charged pion or kaon ($\ell - h(n\pi^{0})$). 
This efficiency rises sharply from
a threshold at $x_{\pm} \sim 0.2$
to 60\% at $x_{\pm} \sim 0.25$,
and reaches a plateau of
about 70\% by $x_{\pm} \sim 0.5$.  Studies of radiative mu-pair 
($\mu\mu\gamma$) events in
both data and simulation indicate a 
rate for misidentification of a muon as a pion of less than 2\% per
track over the kinematic range of this analysis.

The resulting 2041 $h^{+}h^{-}$ candidates
are binned in $x_{+}~vs.~x_{-}$ with bin size 0.1
and range $0.2$$<$$x_{\pm}$$<$$0.95$,
as shown in 
Fig.~\ref{fig:dataxpxm}.  
The missing corner bins at $x_{+}$=$x_{-}$ are the result
of the energy criteria that help suppress $\gamma\gamma$
events and QED events.
This figure clearly shows the depletion in the other corners that have
one ``hard'' and one ``soft'' hadron, as expected for $V \pm A$.
In addition we examine the product $c_{+}c_{-}$ as shown in 
Fig.~\ref{fig:datacpcm}.

The Monte Carlo simulations start with the KORALB~\cite{KoralB}
tau-pair generator and use the GEANT package~\cite{GEANT}
to model the detector response. 
Monte Carlo studies indicate that over
97\% of these events 
were of the five final states
$\pi^{+}\pi^{-}\nu\bar{\nu}$ (81\%), 
$\pi^{\pm}K^{\mp}\nu\bar{\nu}$ (9\%),
$\pi^{\pm}\rho^{\mp}\nu\bar{\nu}$ (4\%), 
$\pi^{\pm}\mu^{\mp}\nu\bar{\nu}(\nu/\bar{\nu})$ (2\%), and 
$\pi^{\pm}K^{*\mp}\nu\bar{\nu}$ (2\%);
the next leading mode was
$\pi^{\pm}e^{\mp}\nu\bar{\nu}(\nu/\bar{\nu})$
(0.8\%).
For these five specific decay channels large samples were generated with
both ($V$$-$$A$) and pure $V$ coupling. In the case of 
$\pi^{\pm}\mu^{\mp}\nu\bar{\nu}(\nu/\bar{\nu})$
events were
also thrown with a ($V$$+$$A$) coupling; for the other final states the 
($V$$+$$A$) and ($V$$-$$A$) predictions are identical.   
These events are then appropriately scaled to be added together for
comparison to the data distributions.
For each bin in either the 
$(x_{+},x_{-})$ or $c_{+}c_{-}$ analysis,
the number of predicted events for {\it any} given value
of $\xi_{h}$ is expressed in terms of the number from the
($V$$-$$A$), ($V$$+$$A$) and $V$ simulations by

\begin{equation}
n_{i}(\xi_{h}) = C_{-1}N^{V-A}_{i} +
                 C_{+1}N^{V+A}_{i} +
                 C_{0}N^{V}_{i} 
\end{equation}
with $C_{\pm 1}$$=$$(\xi^{2}_{h} \pm \xi_{h})/2$ 
and $C_{0}$$=$$(1 - \xi^{2}_{h})$.
A binned maximum likelihood fit 
in the one parameter, $\xi_{h}$, 
is then performed, 
with the resulting likelihood distribution 
for the $(x_{+},x_{-})$ analysis
shown in Fig.~\ref{fig:likexpxm}.
The result is 
$\vert\xi_{h}\vert$ = $0.998^{+0.063}_{-0.059}$, 
with the uncertainty being purely statistical.  A similar
likelihood curve is found for the $c_{+}c_{-}$ analysis~\cite{0D}, 
yielding $\vert\xi_{h}\vert$~ = $1.011^{+0.064}_{-0.060}$. 

We have made many systematic checks which can be divided into four
broad categories: 
possible biases in the methodology, 
our estimate of backgrounds,
dependence on the 
simulation
of the variables used in event selection,
and possible biases in the simulations.  
These systematic effects have been studied for both the 
$(x_{+},x_{-})$ and $c_{+}c_{-}$ analyses. 
Because the values of $\vert\xi_{h}\vert$ from the two
analyses are statistically indistinguishable,
only the details for
the $(x_{+},x_{-})$ analysis will be given.

In addition to the likelihood technique, a $\chi^{2}$ analysis was
performed, with 
minimal shift in the value of $\vert\xi_{h}\vert$.  
Further, studies with very large
samples of simulated events 
(which, for speed, had their kinematic variables smeared
with typical resolution functions instead 
of full detector simulation)
showed that the 
choice of the likelihood estimator causes no inherent shift in the 
value of $\vert\xi_{h}\vert$. 
Various choices of bin size were tried for both the
likelihood and $\chi^{2}$ analyses, from which we assign a systematic
uncertainty of $\pm$0.02.

Studies with Monte Carlo samples indicate a bias of $-0.011 \pm 0.006$
in the fitting procedure and of $-0.020 \pm 0.010$ due to the use of
only the five most common modes.  We include these shifts to obtain a
central value of $\vert\xi_{h}\vert$~$=1.029$ and incorporate the 
uncertainties on the shifts into the overall systematic error.

If the Monte Carlo were not 
properly handling the various $\tau^+\tau^-$ decays
or if there were non-trivial amounts of non-tau backgrounds, then
changing the tau branching fractions 
would affect the result because the various final states have 
different distributions in the $(x_{+},x_{-})$ plane.  Variations of 10-20\%
in these 
branching fractions produced changes in $\vert\xi_{h}\vert$ of less than 0.01.
We used
an extended likelihood technique to incorporate the normalization
into the fit, which yielded a tau branching fraction
and statistical uncertainty 
of ${\cal B}_{\pi}$ 
= $0.11 \pm 0.01$, consistent
with the established value~\cite{BKHPRD,PDG}.
This extension of the likelihood to include normalization
had a negligible effect on the
fitted value of $\vert\xi_{h}\vert$,
changing it by 0.01.

The effect of possible contamination from the $\mu\mu\gamma$
final state was investigated by introducing simulated
$\mu\mu(\gamma)$ events into the data sample.  After weighting
events that passed all but the muon rejection criteria by roughly
five times the rate at which muons fake hadrons, a shift in
$\xi_{h}$ of $-$0.02 is observed.  The resulting uncertainty on
$\vert\xi_{h}\vert$, assuming the actual measured 
misidentification
rate, is less
than 0.005.
 
The Monte Carlo simulations for 
the final state $\pi^{+}\pi^{-}\nu\bar{\nu}$
were examined to see if any of the efficiencies of the various 
selection criteria were different for $V$ and $V$$-$$A$.   
Of all the selection criteria, the restriction on 
$\sin\Theta_{{\rm min}}$ is the one to which $\xi_{h}$ is most
sensitive, especially because it tends to eliminate events
along the diagonal in the $(x_{+},x_{-})$ plane.
We therefore
varied the
limit on ${\rm sin}\Theta_{{\rm min}}$
down to 0.05
and up to 0.20, observing changes in the fitted value of 
$<$$0.02$.
Replacing the limit on ${\rm sin}\Theta_{{\rm min}}$ with requirements
on acoplanarity or visible event energy (also effective
in suppressing $\gamma\gamma$ backgrounds, but less so) 
produced similar changes in $\vert\xi_{h}\vert$, but of the opposite sign.
An uncertainty of $\pm 0.01$ was assigned to this effect.

Other event selection variables were also evaluated by varying the 
requirements, leading to very small uncertainties.  An exception
was the $h/\mu$ discrimination, for which we tried using
various subsets of the four criteria as well as changing the limits
in them individually, leading to an associated
uncertainty of $\pm 0.025$. Similarly we tightened and loosened (where 
possible) the trigger requirements with effects observed of
$\pm 0.015$.

All of the efficiencies depend on the Monte Carlo simulations.  
The most critical aspects of this are the photon veto
(for which hadronic splitoffs need to be properly modeled), the
$h/\mu$ discrimination, and the trigger.  
To investigate these
more thoroughly, samples of events were obtained in both data and
in Monte Carlo simulation for 
the $\ell$-$h(n\pi^{0})$ topology previously described.
From these studies we obtained
$x_{\pm}$-dependent corrections to the efficiencies for all three
aspects of the analysis.  Applying these indicated very small
shifts, some positive and some negative, in the fitted value of 
$\vert\xi_{h}\vert$.  
Based on this we applied no shift in the central value and
assigned a systematic uncertainty of $\pm 0.01$.

Taking
all the effects in quadrature yields an overall systematic
uncertainty of 
$\Delta\xi_{h}$~ = $\pm 0.040$ for the $(x_{+},x_{-})$ analysis.
For the $c_{+}c_{-}$ analysis, the biases shift the central value to
$\vert\xi_{h}\vert$~$=1.036$ with a systematic uncertainty of 
$\Delta\xi_{h}$~ = $\pm 0.049$.  Our quoted value for the helicity
and its uncertainties is the simple average of these two analyses.

In summary, we have measured the 
magnitude of the helicity of the tau neutrino
from the energy-energy correlation in
$\tau^+\tau^-$$\rightarrow$$(h^+\overline{\nu})(h^-\nu)$ 
events,
obtaining
$\vert\xi_{h}\vert$$=$1.03$\pm$0.06$\pm$0.04. 
This is consistent with the other measures of the $\nu_{\tau}$
helicity~\cite{ARGUS1,ARGUS2,ALEPH,L3,OPAL}.  
Considering only $V$ and $A$ interactions, the physical
bounds are 0$<$$\vert\xi_{h}\vert$$<$1, and our result
corresponds to $\vert\xi_{h}\vert$$>$0.87 at 95\%CL.
From Eqn. 5 this implies at most 24\% pure $V$ (or pure $A$) at 95\%~CL.

We gratefully acknowledge the effort of the CESR staff in providing us with
excellent luminosity and running conditions.
This work was supported by 
the National Science Foundation,
the U.S. Department of Energy,
the Heisenberg Foundation,  
the Alexander von Humboldt Stiftung,
Research Corporation,
the Natural Sciences and Engineering Research Council of Canada,
and the A.P. Sloan Foundation.

\newpage 
%
%

%
%
\newpage

\simplex 
{t}
{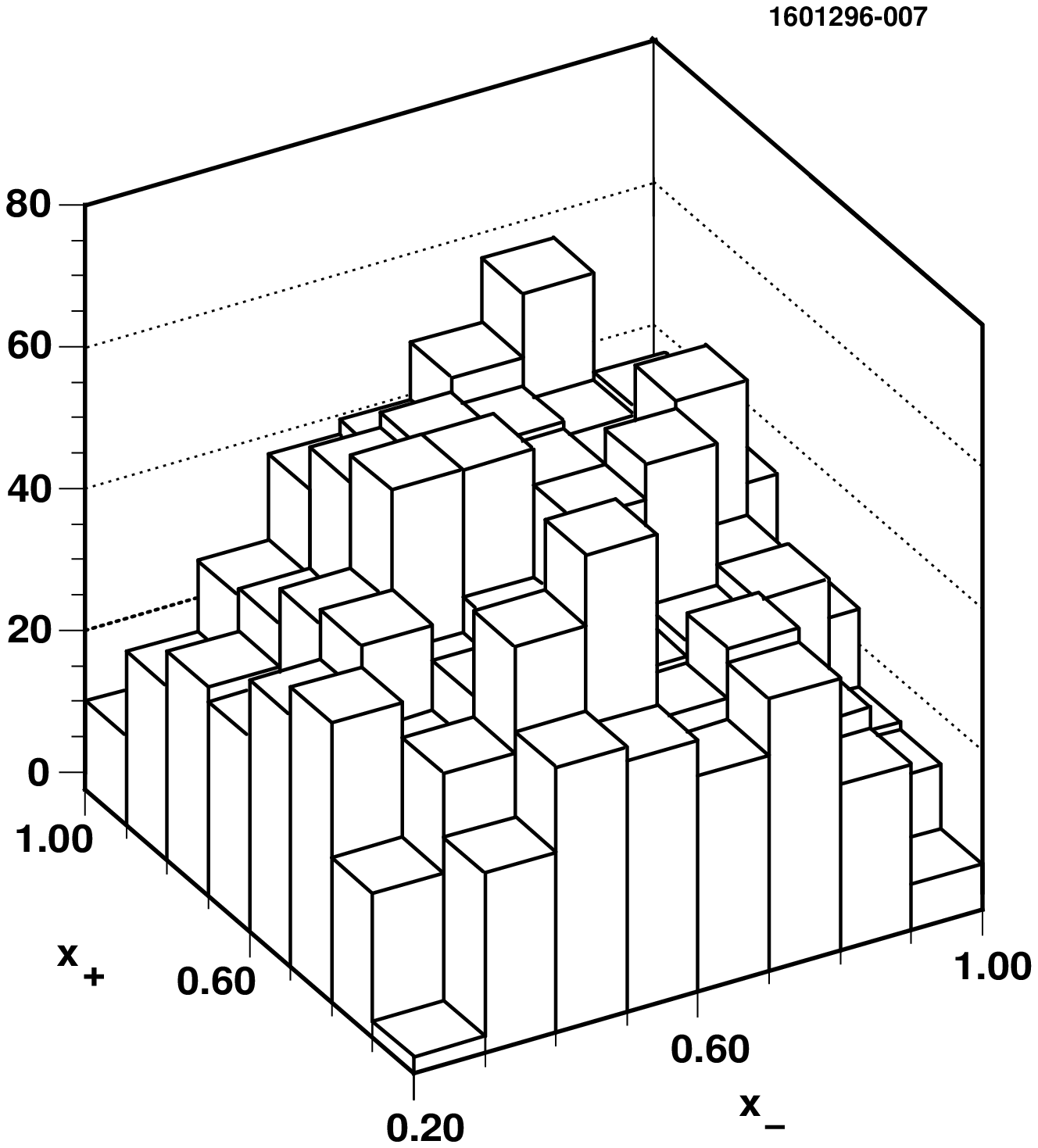}
{dataxpxm} 
{ {The energy-energy distribution for the data, uncorrected for 
efficiencies, using the scaled energy variables.  Note that the 
actual range of the data is 0.2$<$$x_{\pm}$$<$0.95. }} 
{\hsize}

\newpage
\simplex 
{t} 
{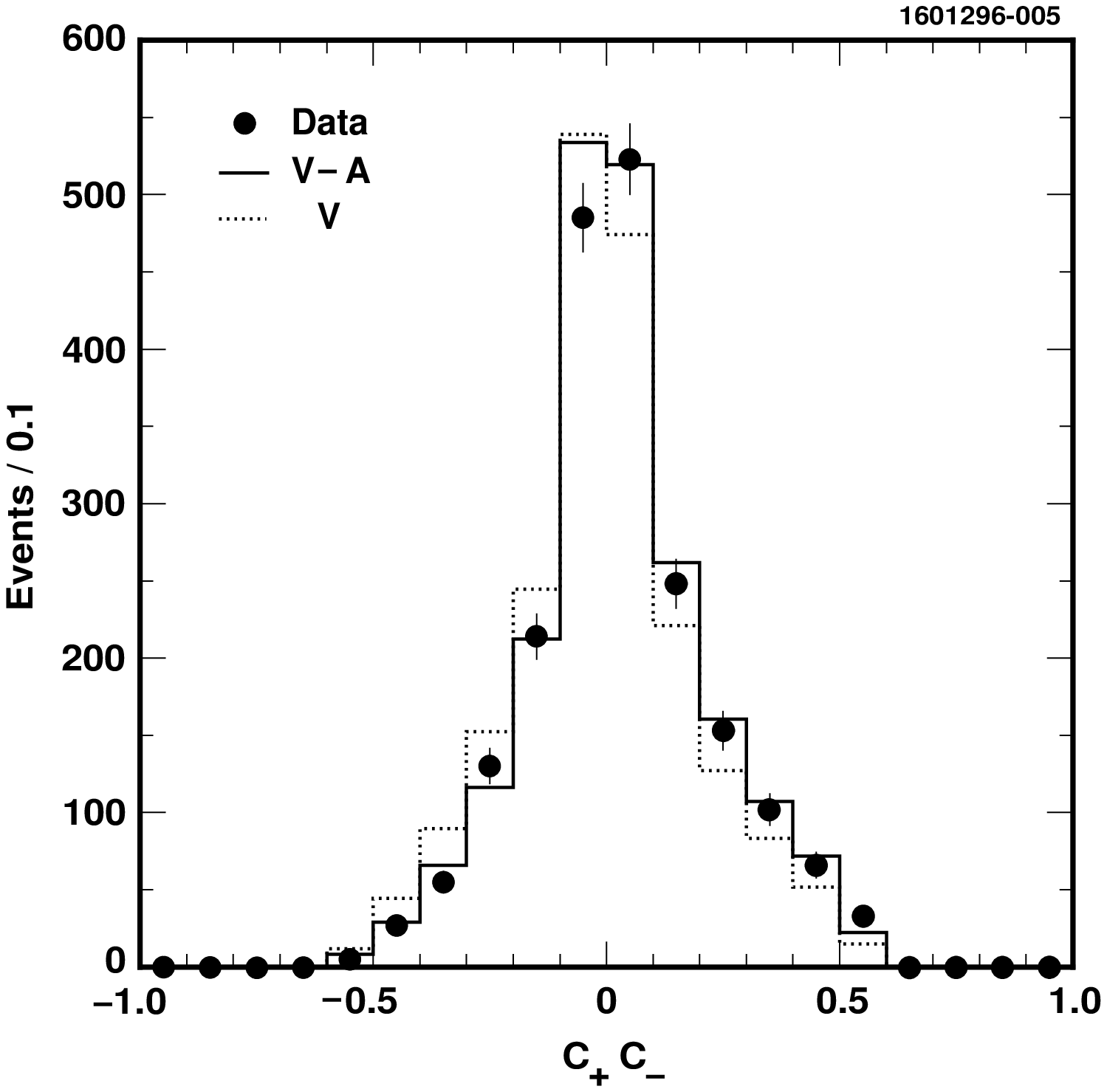} 
{datacpcm} 
{ {The distributions of $c_{+}c_{-}$ for data and for $V$$-$$A$ 
and $V$ simulations. } } 
{\hsize}

\newpage
\simplex 
{t} 
{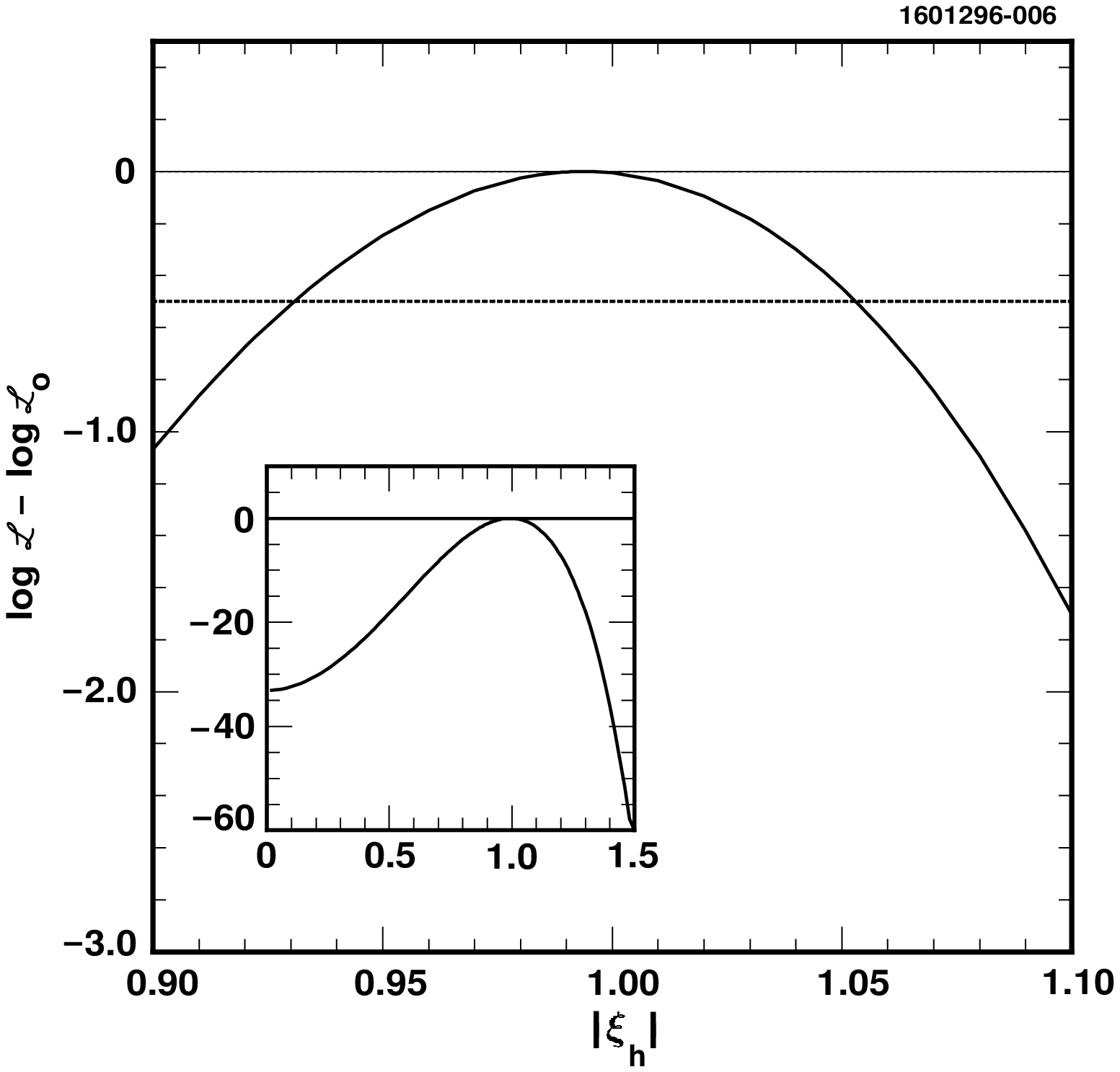} 
{likexpxm} 
{ {Log likelihood $vs.$ $\vert\xi_{h}\vert$ for the $(x_{+},x_{-}$) 
analysis.  The maximum log likelihood value has been subtracted.
The insert shows the full range explored.} } 
{\hsize}


\begin{thebibliography}{99}
\bibitem{0D}  Using the mean of this distribution, 
i.e., $<$$c_{+}c_{-}$$>$,
as the estimator gives $\vert\xi_{h}\vert$=$1.05\pm0.06$.  No 
systematic study of this zero-dimensional analysis was performed.
\end{thebibliography}
\end{document}